# AI-Driven Document Redaction in UK Public Authorities: Implementation Gaps, Regulatory Challenges, and the Human Oversight Imperative

Yijun Chen, Macquarie University,NSW, Australia

Email: yijun.chen4@students.mq.edu.au

Document redaction in public authorities faces critical challenges as traditional manual approaches struggle to balance growing transparency demands with increasingly stringent data protection requirements. This study investigates the implementation of AI-driven document redaction within UK public authorities through an innovative methodological approach utilizing Freedom of Information (FOI) requests. While AI technologies offer potential solutions to redaction challenges, their actual implementation within public sector organizations remains critically underexplored. Through strategic sampling of 44 public authorities across healthcare, government, and higher education sectors, this research reveals significant gaps between technological possibilities and organizational realities. Findings demonstrate remarkably limited AI adoption (only one authority actively using AI tools), widespread absence of formal redaction policies (50% reporting "information not held"), and deficiencies in staff training. The study identifies three key barriers to effective AI implementation: poor recordkeeping practices, lack of standardized redaction guidelines, and insufficient specialized training for human oversight. These findings have significant implications for regulatory frameworks and operational practices, highlighting the need for a socio-technical approach that balances technological automation with meaningful human expertise. This research contributes the first empirical assessment of AI redaction practices in UK public authorities, establishing an evidence base for policymakers navigating the complex interplay between transparency obligations, data protection requirements, and emerging technologies in public administration.

## 1 INTRODUCTION

Citizens in democratic countries have an extensive right to access and request information held by contemporary public-sector organizations [1]. The responses to these requests must be reviewed for document relevancy, and any sensitive or personal information must be redacted before releasing it to the public. Traditionally, document redaction has been manual, relying on human reviewers to identify and obscure sensitive data such as names, addresses, and classified details [2, 3]. However, the escalating volume of FOI requests has placed unprecedented pressure on public authorities, exposing the limitations of these manual processes. In 2023, UK authorities received 70,475 FOI requests, marking a 34% increase from 2022, with fewer timely responses [4]. This dramatic surge in demand, coupled with a declining ability to meet statutory deadlines, underscores a critical challenge for current manual redaction processes while heightening the risk of errors and delays that jeopardize both transparency and privacy in individuals and organizations.

Historically, several high-profile failures underscore the consequences of inadequate redaction, including the National Security Agency scandal [5] and the AstraZeneca Contract case [6]. More recently, a 2023 High Court ruling in R (IAB & Ors.) v Secretary of State for the Home Department & Anor [2023] EWHC 2930 (Admin) criticized inconsistent redaction practices, while a UK Information Commissioner's Office (ICO) reprimand highlighted systemic lapses leading unauthorized disclosures [7]. These incidents illuminate a broader crisis wherein increasing transparency demands conflict with stringent privacy obligations under frameworks such as the General Data Protection Regulation (GDPR).

In response, researchers and practitioners have turned to automated techniques for document redaction [8]. Early studies have shown the promise of using Artificial Intelligence (AI) algorithms trained to recognize patterns and identify personal and sensitive information in documents, such as names, addresses, social security numbers, and other personally identifiable data [9]. By automating the initial stages of redaction, AI can improve efficiency, reduce the time required for manual review, and potentially enhance the consistency of redaction decisions [10]. Despite progress in AI-driven redaction, significant gaps remain in understanding its real-world implementation in public authorities. First, research often focuses on the technical design of AI redaction tools, neglecting the implementation phase, where many practical challenges and implications emerge [11]. This includes understanding how these AI tools affect existing redaction workflows, policies, and the roles of human staff within public authorities. Crucially, the impact of AI is highly context-dependent, requiring consideration of existing institutional structures, processes [12], and social and political contexts [13, 14]. Second, successful implementation requires specific expertise – not just in operating the AI tools but in understanding their limitations, interpreting their outputs, and reconciling them with existing legal and ethical requirements [15]. This argument provides a rational explanation for why existing legal frameworks, including Article 22 of the GDPR, explicitly require human involvement as a "**default setting**" in these systems, thereby ensuring accountability in automated processes [16].

However, significant deficiencies remain in comprehending the human aspect of this oversight: How is staff instructed to assess existing redaction methods and AI-assisted redactions? What qualifications guarantee their proficiency in reconciling automated outputs with manual redaction? The alignment, or misalignment, between AI systems and current manual workflows, is equally underexplored. This gap undermines the evaluation of both the operational effectiveness of automated redaction technologies and their adherence to legal frameworks such as the GDPR, which relies on stringent human accountability. To address these critical gaps, this study aims to investigate the current state of AI-driven document redaction within UK public authorities. Unlike prior studies that rely on self-reported data or policy analysis, this research is the first to utilize Freedom of Information (FOI) requests to gain unfiltered access to the operational realities of AI redaction within these organizations. This approach provides a level of detail and authenticity that is difficult to achieve through other research methods. Given the exploratory nature of this research and the novel use of FOI requests, this study focuses on a carefully selected sample of 44 public authorities across the UK government, education, and healthcare sectors. While this sample size covers a range of representatives, it allows for an in-depth analysis of the complexities of AI adoption in this specific context. The findings will provide valuable insights that can inform the design of future research involving larger, more representative samples. To achieve these aims, the study is guided by the following core research questions:

**RQ1:** What formal policies or standards guide redaction practices in UK public authorities? How do public organizations integrate AI and automated systems into their document redaction processes, and what role do human reviewers play in safeguarding information to compliance with data protection standards?

**RQ2:** What are the key challenges faced by public organizations when using AI-driven document redaction tools, and how are staff trained to effectively identify and handle sensitive information?

By addressing these questions in an exploratory manner, this study endeavours to provide a comprehensive understanding of both the policy landscape and implementation influencing document redaction in the public sector. Furthermore, by leveraging Freedom of Information (FOI) requests, this research moves beyond theoretical policy frameworks to obtain first-hand insights into the actual practices employed by these industries. This study provides, for the first time, an empirical assessment of how AI tools are actually being used for redaction in UK public authorities, shedding light on the interplay between policy, technology, and human oversight in this sensitive domain. These advancements are essential not only for enhancing operational efficiency and minimizing human error but also for maintaining the integrity and confidentiality of sensitive information in an increasingly digital and data-driven governmental landscape.

## 2 LITERATURE REVIEW

### 2.1 The Fundamental Challenges of Document Redaction in Public Authorities

Public authorities face a critical dilemma: they must simultaneously uphold the democratic right of citizens to access government information [1, 17] while protecting sensitive personal data in accordance with current data protection law [18]. This tension, far from being merely operational, represents a fundamental challenge to modern governance. Current redaction approaches, however, are increasingly inadequate for addressing this balance. Prior studies have demonstrated traditional document redaction in public authorities relies overwhelmingly on manual processes that are becoming unsustainable in the digital age [10]. While public records collections continue to expand exponent, **specific**ally ranging from emails to official reports (Phiri, 2016), the human-centred processes used to review and redact them have remained largely unchanged [19]. This creates what can be characterized as a "redaction bottleneck" in transparency workflows, where the volume of information requiring processing far exceeds institutional capacity.

The consequences of this bottleneck are profound and measurable. The dramatic 34% increase in FOI requests between 2022 and 2023, coupled with a 5% decline in timely responses [4], is not merely a statistical anomaly but evidence of a structural inconsistency in information governance systems [20] and legal framework [21]. For the purposes of clarity, the core terminology used in this article is summarized in Table 1, including the respective roles of each actor. This background states the law as it was on the 1st of December 2024. In this case, a relevant principle such as Section 14 of FOIA (2000) in the UK allows refusal if a request is considered vexatious or too broad, often because the excessive workload due to redaction makes it difficult to comply [22]. Legal precedents have established redaction burden as grounds for request refusal. In *Salford City Council v ICO and Turkey Accounts Ltd* (EA/2012/0047, 2012), the council successfully argued that excessive redaction requirements justified refusal under Section 14(1) of FOIA. The Upper Tribunal reinforced this position in *Home Office v Information Commissioner and Cruelty-Free International* [2019] UKUT 299 (AAC), confirming that substantial redaction work constitutes a legitimate **"burden"** for invoking S.14 of FOIA (2000). However, this creates a significant regulatory gap: while authorities can refuse requests based on redaction burden, the considerable time spent identifying potentially sensitive information before actual redaction occurs is not explicitly factored into the cost-based refusal threshold under s.12 of FOIA (2000). This legislative inconsistency places public authorities in an untenable position where they cannot fulfil their statutory obligations under FOIA due to resource constraints. The consequence is a systemic risk of either non-compliance or selective compliance with information requests, fundamentally undermining the principles of transparency and democratic access to public information that FOIA was designed to uphold.

Table 1: Summary of Terminology with References This include widely used abbreviations, where appropriate

| Term (and abbreviation) | Summary | Source |
|---|---|---|
| Freedom of Information Act (2000) ("FOIA (2000)") | The primary legislation governing the public's right of access to information held by public authorities in the UK, subject to specified exemptions. | [23] |
| General Data Protection Regulation ("GDPR") | European Union regulation governing the processing and protection of personal data, emphasizing data accuracy, minimization, and accountability, with specific provisions on automated decision-making. | [24] |
| Information Commissioner's Office ("ICO") | The UK authority responsible for overseeing compliance with FOIA and GDPR, providing guidelines and managing complaints regarding information rights. | [21, 22] |
| Freedom of Information request ("FOI request" or "information request") | A formal request made to a public authority to access recorded information, governed by the FOIA (2000). | [20, 25] |
| Subject Access Request ("SAR") | A formal request submitted by individuals to access personal information that an organization holds about them, as mandated under data protection laws such as GDPR. | [26] |
| Appropriate Limit (the "cost limit" or "s.12 FOIA (2000)") | Defined by FOIA (2000), this refers to the statutory limit on the time or cost public authorities are expected to spend on processing information requests. | [21] |
| Section 16 of FOIA (2000) ("s.16 FOIA (2000)") | A statutory duty requiring public authorities to provide advice and assistance to requesters to help them effectively frame FOI requests. | [27] |
| Section 14 of FOIA (2000) ("s.14 FOIA (2000)") | Allows public authorities to refuse FOI requests if they are considered vexatious or overly burdensome. | Cherry and McMenemy [28] |
| Section 10(1) of FOIA (2000) ("s.10(1) FOIA (2000)") | Requires public authorities to respond to information requests promptly and no later than 20 working days from receipt. | [29] |
| Article 22 of the GDPR | Grants individuals' specific rights related to automated decision-making processes, requiring human oversight and the right to contest decisions made solely by automation. | [30] |
| Article 14(1) of the EU Artificial Intelligence Act ("AIA") | Mandates human oversight for AI systems considered "high-risk," ensuring transparency, accountability, and compliance with ethical standards in AI applications. | [31, 32] |
| Human-in-the-loop | A hybrid approach integrating human oversight with automated systems to ensure accuracy, ethical compliance, and risk management in automated decision-making processes. | [33] |

Moreover, current approaches to redaction suffer from systematic technological vulnerabilities that cannot be attributed to simple human error [34]. Recent research by Chen and Kirkham [25] demonstrated convincingly that inadequate documentation of redaction policies creates systemic risks that extend beyond individual mistakes. The redaction failures in high-profile cases like Wikileaks [35], the Mueller Report [36], and Brexit negotiations [37] exemplify how seemingly minor technical oversight such as hidden metadata [34], ineffective PDF tools, or simple marker pen redactions—can lead to significant security and diplomatic consequences [38]. What these cases collectively demonstrate is not isolated incompetence but rather the fundamental inadequacy of current approaches in an increasingly complex digital environment [25]. This argument directly challenges the prevailing assumption that existing redaction practices simply need incremental improvement. Instead, the evidence suggests that traditional approaches have reached their effective limits, creating both a crisis in transparency and unacceptable risks to privacy. This critical assessment of the current state establishes the foundational need for exploring AI-driven alternatives that might resolve these fundamental tensions.

## 2.2 AI Adoption: Promise and Governance Gaps

*2.2.1 The Paradox of AI Governance in UK Public Administration*

AI adoption in public administration has gained momentum across healthcare, welfare, policing, and administrative services, promising enhanced efficiency and data-driven decision-making [39-41]. Practical examples include predictive analytics for optimizing healthcare resources, AI-driven risk assessments in welfare allocation, and chatbots handling routine public inquiries [42]. Despite these advances, AI governance in the UK remains fragmented. Rather than establishing dedicated AI regulatory frameworks, public administration defaulted to relying on existing legislation—primarily the UK GDPR and administrative law—which were not designed to address the unique challenges posed by algorithmic decision-making [43]. As such, Bullock, Chen, Himmelreich, Hudson, Korinek, Young and Zhang [44] argues that this reliance on sector-specific bodies such as the ICO creates regulatory blind spots where accountability mechanisms fail to match the pace of technological deployment. In response, the 2023 AI White Paper's five broad principles: safety, transparency, fairness, accountability, and contestability, while laudable, lack statutory enforcement mechanisms, effectively creating what Zuiderwijk, Chen and Salem [45] characterizes as "**soft governance**" that fails to provide clear guidance for public sector implementation. This critique is particularly relevant for AI-driven redaction, which operates at the intersection of transparency requirements and data protection obligations—an area where soft governance is insufficient to resolve fundamental conflicts between competing values.

While AI applications in policing and welfare have attracted significant public scrutiny and academic analysis [39, 40], the growing use of AI-powered redaction tools has received comparatively little attention. This oversight is problematic because redaction decisions, unlike many other automated processes, directly impact the civic right to government information. The lack of specific guidance for automated redaction creates what Roberts, Babuta, Morley, Thomas, Taddeo and Floridi [43] identify as an accountability gap, where neither existing FOI frameworks nor AI governance principles adequately address the unique challenges of algorithmic transparency decisions. This governance vacuum raises fundamental questions about whether current regulatory approaches can effectively manage the risks of AI-driven redaction while enabling its potential benefits. As Mellouli, Janssen and Ojo [46] argue, the empirical evidence on AI implementation in public governance remains sparse compared to the rapid pace of adoption, creating a dangerous knowledge gap that this study aims to address.

*2.2.2 Critical Assessment of AI Approaches to Redaction*

The evolution of AI technologies for document redaction reveals significant theoretical advancements but also exposes persistent limitations that challenge their practical implementation in public authorities [47]. A critical analysis of this development demonstrates that technical capabilities alone are insufficient without addressing the organizational and ethical frameworks necessary for responsible deployment. In the history of automated redaction research, early rule-based approaches to automated redaction, such as those developed by Cumby and Ghani [8] and Abril, Navarro-Arribas and Torra [48], were fundamentally constrained by their reliance on static ontologies like WordNet. These systems operated on a flawed assumption that sensitivity is an inherent property of entity types rather than contextually determined—a simplification that fails to capture the complex nature of information sensitivity in public documents. Li, Sun, Han and Li [49] effectively critique this approach, demonstrating that entity-based redaction produces both false positives (over-redaction) and false negatives (under-redaction) because sensitivity is inherently contextual rather than categorical. In addition, more sophisticated probabilistic approaches, such as Sánchez et al's [50] Information Content (IC) method, attempted to address this limitation by identifying sensitive phrases based on statistical uniqueness. However, Wang, Tan,

Yang, Yuan, Wang, Chen, Ren, Zhang and Shao [51] expose significant implementation barriers in these approaches—notably computational overhead and scalability challenges—that make them impractical for large-scale deployment in resource-constrained public authorities. This critique revealed current a disconnect between laboratory experimentation and operational realities, a gap that remains insufficiently addressed in technical literature.

Recent advances in zero-shot text sanitization (ZSTS) by Albanese, Ciolek and D'Ippolito [52] and end-to-end systems like iDox.ai described by claim impressive performance metrics (98% recall and 97.1% accuracy respectively) [10]. However, these evaluations contain a critical methodological flaw: they typically measure performance against pre-defined datasets rather than in complex, real-world governance contexts where sensitivity is often ambiguous and contested. This fundamental gap between controlled evaluation and authentic implementation—requires closer examination to determine the true potential of these technologies in public administration [53]. The competing claims in the literature expose a theoretical tension between optimistic technical perspectives [10, 52] and more skeptical implementation studies [54] that found automated tools achieving only 67% recall in real-world clinical settings. This contradiction suggests that technical capabilities alone cannot resolve the complex challenges of automated redaction without addressing the organizational, ethical, and governance frameworks necessary for their responsible deployment.

*2.2.3 The Critical Necessity of Human Oversight: Ethical and Legal Imperatives*

The limitations of AI-driven redaction systems reveal not merely technical challenges but fundamental ethical and legal imperatives for maintaining meaningful human oversight. This necessity stems from three interrelated factors: vulnerability to deanonymization, algorithmic bias, and legal requirements that collectively establish human review as non-negotiable rather than merely desirable. For example, Beltrame, Conti, Guglielmin, Marchiori and Orazi [55] proposed the development of RedactBuster—capable of inferring redacted entity types with 98.5% accuracy even when text is completely removed—fundamentally challenges the assumption that automated redaction can fully protect sensitive information. This work demonstrates that contextual information surrounding redactions can leak sensitive data regardless of the technical sophistication of the redaction method. This vulnerability is not merely a technical limitation but reveals a fundamental epistemological problem: the meaning of text exists not only in specific terms but in their relationships and contexts, which AI systems struggle to fully comprehend and protect.

Beyond technical vulnerabilities, Mansfield, Paullada and Howell [56] expose a troubling ethical dimension: commercially available and open-source PII masking systems exhibit significant disparities in name detection based on race and ethnicity, with notably higher error rates for names associated with Black and Asian/Pacific Islander individuals. This finding directly contradicts the assumption that automation inherently produces more consistent outcomes than human reviewers. Instead, it reveals how automated systems can systematically perpetuate and amplify existing social inequalities—a particularly serious concern in public authorities obligated to serve all citizens equitably. The practical limitations of current technologies are further highlighted by El-Hayek et al.'s [54] evaluation of four open-source de-identification tools on Australian general practice patient notes. Their conclusion that none were suitable for immediate use, with the best tool achieving only 67% recall, underscores the significant gap between laboratory performance and real-world application across different domains and documentation styles. This disparity raises serious questions about the readiness of current AI redaction tools for deployment in public authorities where high accuracy standards are essential. These empirical limitations acquire legal significance when considered alongside regulatory frameworks. Article 22 of the GDPR specifically restricts automated decision-making with significant effects on individuals [57], while Article 14 of the proposed EU AI Act explicitly mandates human oversight for "high-risk" AI systems [58]. While redaction may not always constitute a "decision" under GDPR Article 22 [58], the consequences of inaccurate redaction—potential disclosure of sensitive personal information—have significant privacy implications that activate these requirements. This legal framework establishes human oversight not as an optional feature but as a core requirement for responsible implementation.

Additionally, another critical gap is that these regulations mandate human oversight without specifying the training and competencies necessary for effective implementation [59]. The absence of standardized training requirements means that personnel often lack the skills to critically evaluate and correct AI outputs, creating what Kirkham [15] identifies as a dangerous knowledge asymmetry between automated systems and their human supervisors. This gap reinforces the argument that comprehensive training addressing both technical skills and ethical considerations is not merely beneficial but essential for the responsible implementation of AI-powered redaction. Therefore, the combined weight of these technical, ethical, and legal arguments establishes that AI-driven redaction cannot function as an autonomous replacement for human judgment but must instead operate within a carefully designed socio-technical system where human expertise and oversight remain central. This conclusion directly informs our research questions about the integration of AI systems into redaction workflows and the critical role of human reviewers in safeguarding sensitive information.

## 3 METHODOLOGY

This study employs a mixed-methods approach to investigate the practices, challenges, and implications of AI-driven document redaction in public authorities. The primary data collection method involves submitting Freedom of Information (FOI) requests, which serve as a specialized form of "survey", granting researchers legally binding and transparent access to official records [20, 60]. The WhatDoTheyKnow platform was employed to facilitate and document these submissions. This online service automates and archives requests and corresponding responses, thereby promoting comprehensive analysis and ensuring transparency throughout the study. The advantage of FOI lies in its ability to provide transparent and legally binding access to information, allowing for the exploration of detailed responses that would otherwise be unobtainable, particularly in the context of AI-related issues and its legal obligations [28, 61]. In particular, compared with other approaches, such as face-to-face interviews or phone calls, FOI is more useful when comparisons are sought between various public authorities, which is the scenario we are concerned with [62]. As noted by Walby and Luscombe [63], Freedom of Information request has become increasingly recognized as an effective tool for qualitative inquiry in transparency, aligning with the goals of this study to explore the complexities of AI-driven redaction in public administration [46, 64, 65].

### 3.1 FOI Request Content and Rationale

The FOI requests sent to the selected authorities were designed to gather specific information about AI-driven document redaction practices. These questions were carefully crafted to align with the study's objectives, focusing on four key areas: AI redaction processes and policy, staff training, and the operational challenges faced by organizations. In order to clarify each decision, Table 2 provides the questions we sent to organizations and justifications.

Table 2: Overview of the Freedom of Information (FOI) questions regarding document redaction processes and AI adoption. The table presents each question's rationale, expected outcomes or insights, and its alignment with the study's specific Research Quest

| FOI Question | Rationale | Expected Outcomes |
| --- | --- | --- |
| 1. Please describe the process and standards used by your organization to redact documents. Specifically, what guidelines, standards, or procedures are followed to protect personal and sensitive information? | This question gathers information on the formal guidelines and procedures followed by public authorities. It provides insights into how authorities protect personal information and | Expected to provide a detailed overview of the operational standards, helping assess the consistency and effectiveness of redaction practices in relation to |

| FOI Question | Rationale | Expected Outcomes |
|---|---|---|
| | sensitive data in compliance with privacy regulations. | legal and regulatory requirements. |
| 2. Does your organization use any tools, software, or automated systems for document redaction? If yes, please specify the types of tools used (e.g., AI-based systems, machine learning models, manual redaction tools). | Investigates the extent to which AI and automated systems are employed for document redaction, uncovering which technologies are applied and their role in the process. | Insights into the range of AI tools used and their integration into the workflow. This will highlight the level of automation in the redaction process. |
| 3. Please provide details about any AI or machine learning technologies used to support document redaction, including the names of any software or vendors. If an automated tool or AI-based system is used for document redaction, does your organization include a human review step in the process before finalizing redaction? Please describe the role of human reviewers. | Explores how AI technologies are integrated into the redaction process, specifically regarding the role of human oversight in ensuring accuracy and compliance. | Understand the balance between automation and human intervention, revealing how public authorities integrate human review to mitigate AI errors. |
| 4. Please provide details on training provided to individuals involved in the document redaction process, particularly for AI-based or automated redaction tools. How are staff trained to identify sensitive information and use these tools effectively? | Examines how staff are trained in using AI tools and how they are equipped to identify and protect sensitive information. | Clarifies the level of expertise and preparedness of staff, essential for understanding the effectiveness of training programs in AI adoption and compliance with data privacy standards. |
| 5. What are the biggest challenges your organization faces when redacting documents, particularly when using automated or AI-based systems? | Identifies operational, technical, and organizational barriers to adopting AI-driven redaction, offering insights into the practical challenges of AI adoption in public sector organizations. | Expected to uncover the real-world challenges public authorities face in using AI, providing insights into areas requiring improvement or additional resources. |

### 3.2 Choice of Respondents

This research focused on public authorities as respondents, drawing on prior literature, including Chen and Kirkham [25], Cherry and McMenemy [28] and Wasike [62], that recognizes Freedom of Information (FOI) requests as a particularly insightful method for examining public institutions and, more broadly, as a valuable tool for studying bureaucratic organizations. Accordingly, we employed a carefully curated sample of 44 public authorities, categorized into three major UK public sectors: healthcare, government bodies, and academic institutions. These sectors were selected because they routinely handle sensitive data and increasingly rely on AI-driven redaction tools. Each sector provides a distinctive context for examining AI adoption, offering a unique perspective on the integration and application of advanced AI systems in public administration. To provide further clarity, Figure 1 illustrates the distribution of these research respondents, and the sections that follow elaborate on each decision-making process.

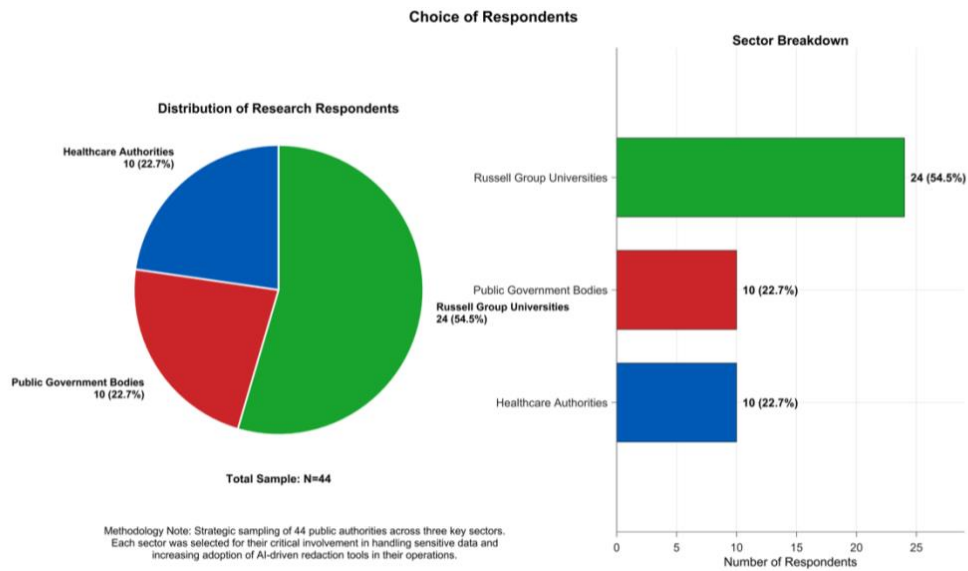

Figure 1: Distribution of the 44 public authorities sampled, illustrating the breakdown across three key sectors—Healthcare Authorities, Public Government Bodies, and Russell Group Universities.

Regarding generalizability, the research sample comprises a strategic selection of key institutional sectors: ten NHS Foundation Trusts known for managing sensitive patient data under GDPR constraints; ten high-level ministerial government bodies that navigate complex legal frameworks while handling confidential governmental information; and twenty-four prestigious Russell Group universities distinguished by their research excellence and early adoption of technology. This deliberately diversified cross-sectoral composition facilitates a comprehensive analysis of AI-based redaction across organizations with distinct regulatory requirements, operational challenges, and institutional objectives, forming a robust foundation for examining how AI technologies can enhance document management and redaction processes in environments where data protection failures could result in significant legal, financial, and reputational consequences.

The exclusion criteria for this study were explicitly centred on the responses of authorities to Freedom of Information requests. In instances where a public authority relied on Section 12 of the FOIA to deny a request, citing that the costs associated with processing the request would surpass the limits outlined in the Act, the request was subsequently revised. We used Section 16 of the FOIA, which requires public authorities to narrow the scope of the request; if, following this modification, the authority continued to fail to respond or declined to provide the requested information, it was subsequently excluded from the study. Furthermore, the rationale for this exclusion lies in the fact that appealing a refusal under Section 12 can be a lengthy procedure, often extending to a duration of a year or more, thereby substantially delaying the data collection process. Consequently, we excluded authorities where the FOI request process experienced delays due to such disputes, thereby ensuring that the study concentrated on authorities capable of providing timely and pertinent data.

### 3.3 Management of FOI Refusals

When we encountered an FOI request refusal, we adhered to a structured approach to minimize disruption to data collection. As outlined by Hazell and Worthy [66], FOI requests may be refused for various reasons. In our case, most public authorities relied on Section 12 of FOIA to refuse our requests, such as the Cabinet Office, the University of Durham, and Cardiff University. As Figure 2 shows, our FOI requests were made in October 2024, and the first step was to review each of the decisions and refusals for compliance with FOI regulations and assess their validity. Following this, we ensured that the public authorities complied with the requirement under the Freedom of Information Act that they must respond to requests promptly and no later than the twentieth working day after receiving the request. In line with this, we adhered to the duty outlined in Section 16(1) of FOIA, which mandates public authorities to provide reasonable advice and assistance to those making FOI requests. We initiated follow-up communication with the relevant organizations to clarify the reasons for the refusal and request partial data where feasible.

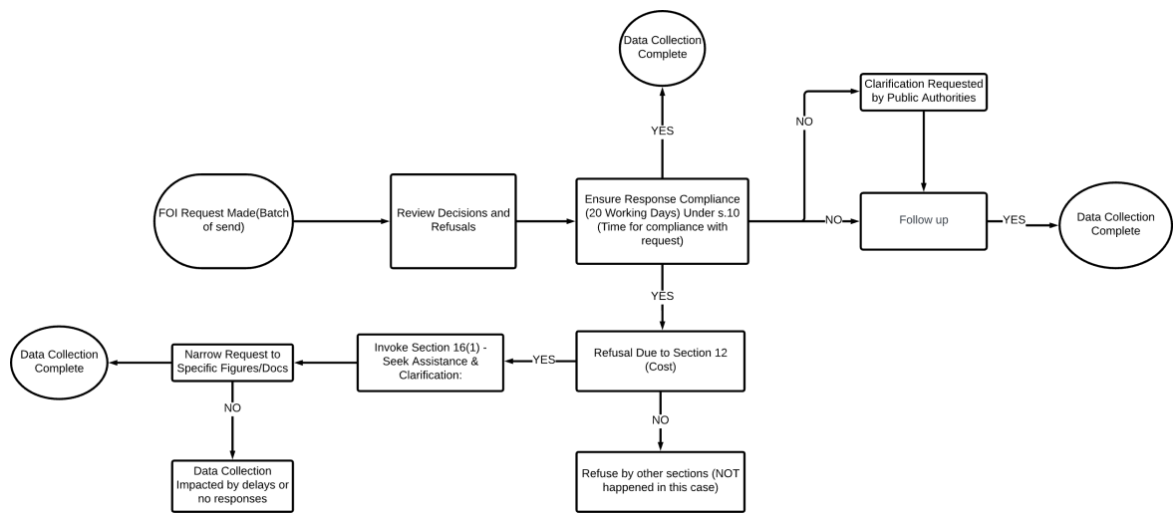

Figure 2: This figure shows the steps taken from initial request submission to documenting the impact of refusals. Key steps include reviewing refusals for compliance, invoking Section 16(1) for assistance, and narrowing the request. The diagram reflects the research process, which did not include internal reviews or appeals to the Information Commissioner's Office.

In response to the refusals, we narrowed our requests to focus on specific documents—such as figures or manuals related to the Freedom of Information Team and the Subject Access Request team—aiming to reduce the cost burden while still ensuring the continued collection of essential data. Once the requests were adjusted, we reiterated that it was possible to fulfil them within the cost limit based on the nature of the documents requested. However, if absolute refusals persisted under Section 12, we escalated the matter by appealing to the Information Commissioner's Office (ICO) for a second review. By maintaining a proactive and professional approach throughout this process, we ensured that the data collection remained robust, minimizing the impact of refusals based on Section 12 on the overall progress of our research.

## 4 DATA ANALYSIS

Based on the FOI responses, we employed a mixed methods approach, conducting two types of analysis. First, we performed a quantitative analysis, presenting summary descriptive statistics for all responding authorities. Excel was used to organize the data and calculate basic statistics, while SPSS facilitated more advanced analysis, including correlations between variables such as authority size and AI adoption. This approach provided a numerical overview of redaction practices across sectors.

Second, we conducted a thematic analysis following Braun and Clarke [67] to code the various redaction policies and responses. This qualitative analysis focused on the specific text corpus we obtained, allowing us to identify patterns and themes in the responses. Thematic analysis was particularly useful due to its flexible approach to handling diverse data, which was essential given the wide range of responses, from brief sentences to highly detailed replies that included existing policy documents. NVivo software was used for efficient coding and theme identification. Text segments were coded to represent specific concepts, such as AI usage and challenges in document redaction, and grouped into broader themes like "AI-driven redaction" and "policy implementation." The analysis was refined to ensure that the identified themes accurately represented the data, supported by excerpts from the responses.

The integration of these methods allowed us to examine both the prevalence of practices (quantitative) and their contextual significance (qualitative). For instance, when analyzing staff training, we quantified how many authorities provided training while also coding training descriptions to understand their nature. In line with Chen and Kirkham [25] and Kirkham [21], our aim was to uncover AI practices and existing challenges that were commonly implemented rather than focusing on isolated concerns or errors. Therefore, this article limits its discussion to issues that were relatively widespread across responses rather than addressing every individual issue that emerged. This integrated analytical framework provided a comprehensive foundation for addressing our research questions about AI-driven redaction in UK public authorities.

## 5 RESULT

This study analyzed redaction practices across UK public authorities using 44 Freedom of Information (FOI) requests distributed across three sectors: Healthcare (10 requests), government (10 requests), and Russell Group Universities (24 requests). As Figure 3 shows, the results indicate that a total of 30 (68.2%) responses were received, while 14 (31.8%) either due to a lack of response or invocation of Section 12 of the FOI Act (cost limits). This non-compliance rate remains high, even with further clarification under Section 16 of the FOIA, which mandates public authorities to provide advice and assistance to requesters to help them make a valid request.

Another notable finding from the results is the significant incidence of "NOT HELD" responses, with 15 of the 30 responses (50%) indicating that the requested information, including redaction policies and AI adoption for redaction, was not available. In detail, "NOT HELD" means that public authorities lack formal guidelines, business processes, or procedures for document redaction. Based on our quantitative analysis, distinct patterns emerged across the three sectors:

- **Healthcare:** Nine out of ten requests (90%) were answered. Among these responses, 4 (44.4%) cited external guidance (primarily from the Information Commissioner's Office (ICO)), 3 (33.3%) relied on internal policies, and 2 (22.2%) responded "NOT HELD."
- **Government:** Four out of ten requests (40%) were answered. All four of these referenced external guidance (from the ICO or the National Archives), and notably none in this sector responded with "NOT HELD."
- **Russell Group Universities:** Seventeen out of 24 requests (70.8%) were answered. However, 13 of these 17 responses (76.5%) were "NOT HELD," indicating no formal redaction policy or procedure. Only two university responses (11.8%) cited ICO guidance, and another 2 (11.8%) referenced internal policies.

Additionally, most of the responding authorities reported relying on general guidelines such as the Freedom of Information Act (FOIA) and Information Commissioner's Office (ICO) regulations. However, neither FOIA nor ICO guidelines explicitly mandate the use of AI-driven tools for redaction. These guidelines focus on compliance with data protection laws, like the General Data Protection Regulation (GDPR), without explicitly specifying how AI should be integrated into redaction practices. This highlights a key gap in current practices. Even when internal policies exist, most organizations lack standard procedures for adopting AI in redaction. For example, the London School of Economics and Political Science stated, "*No standards are used. The Freedom of Information Act and data protection legislation provide the guidelines, meaning that where an exemption applies, the information relevant to that exemption is redacted from the document before release.*" Similarly, Queen Mary University of London said, "*We do not have a process or standard specifically and explicitly relating to redaction, but we follow the ICO guidelines for protecting personal information or other information requiring redaction.*"

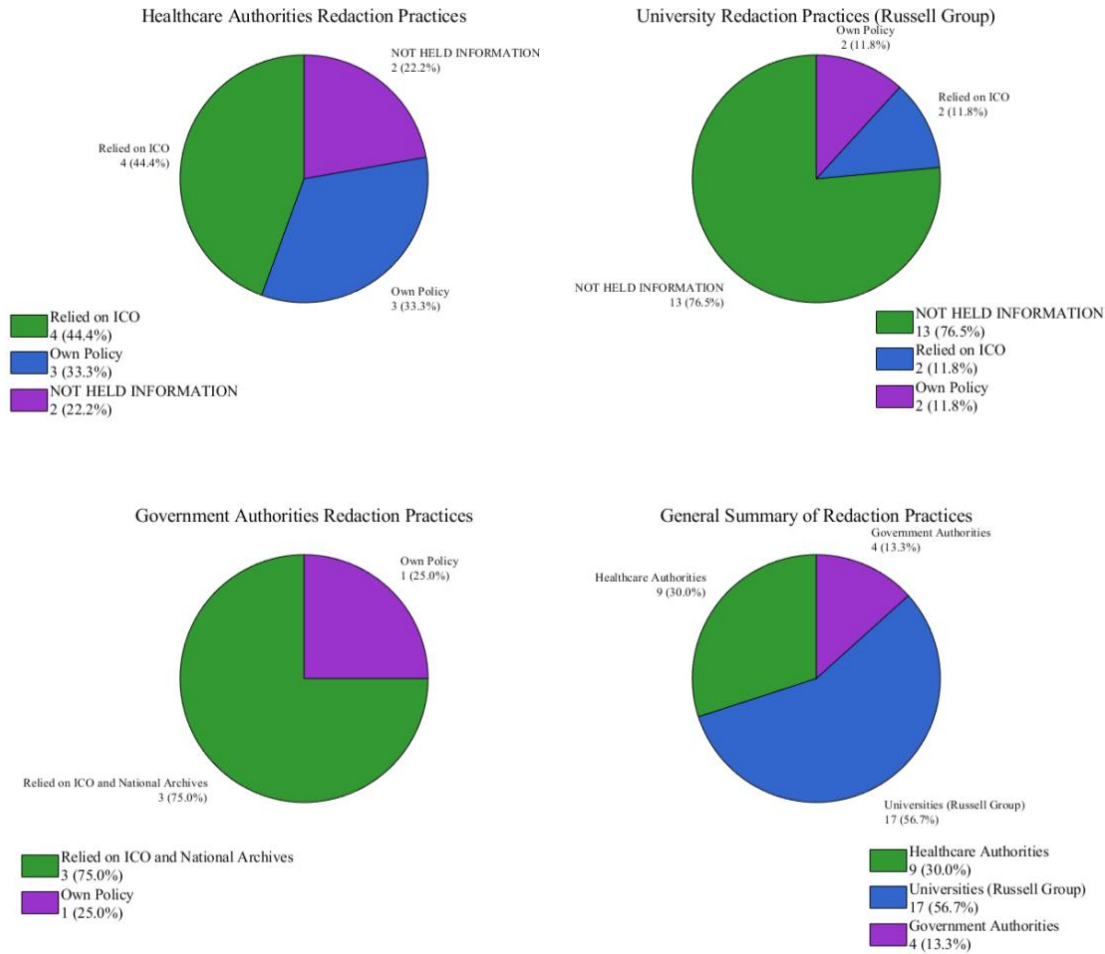

Figure 3: Overview of FOI responses across public authorities.

## 5.1 AI/ML Adoption

The adoption of AI or machine learning tools for document redaction was found to be very limited across the sampled public authorities. Only one authority (out of the 30 that responded) reported actively using an AI-driven tool in its redaction process. This organization is piloting an AI platform called Smartbox.ai to assist in identifying sensitive information during initial document assessments. According to this authority:

We are currently implementing Smartbox.ai, which uses AI to carry out initial assessments of documents to identify personal information. The next step, once completed, is for a team member to review the documents and the proposed redactions made by the software.

In this approach, the AI quickly flags potential personal or sensitive data, and a human reviewer then verifies and finalizes the redactions. This method attempts to balance the efficiency gains from AI with the necessity of human oversight for accuracy. Nonetheless, challenges persist in fully integrating AI, particularly regarding staff training and qualifications for using these tools (as discussed in the next section).

### 5.2 Training and Competence

In this question, we received responses from 6 (20%) public authorities that indicated some form of training, while the remaining authorities either claimed that they do not offer any training, including training on AI-driven tools (7(23%) authorities) or stated that the relevant information was not held (17(57%) authorities). The response "information not held" means a lack of documented training, which leads to the concerning finding that only six authorities have any training in place. Among those authorities that reported having training, the nature and scope of such training varied considerably.

In specific, several responses merely acknowledged the existence of training without providing further details on its content or the individuals involved. For instance, the University of Nottingham stated, "*The University does not use specific training related to redaction. Identifying sensitive information is covered in the mandatory learning. Redaction of documentation for Information Compliance purposes is undertaken by the Information Compliance Team.*" In addition, we noted that a range of delivery methods was used, with the majority mentioning "*in-house*" or "*on-the-job*" training. For example, East London NHS Foundation Trust commented, "*The Trust provides in-house training and external courses where necessary.*" However, it should be noted that they do not utilize automated or AI-driven tools for redaction. The descriptions of training appear to be informal and may lack standardization. In particular, the phrase "retrain where necessary" suggests a reactive approach to training rather than a proactive, structured curriculum. As such, this raises concerns that staff may not receive comprehensive, foundational training at the outset, which could result in inconsistencies in the application of redaction procedures across authorities. Such variability in training could have significant implications for the accuracy and reliability of redaction practices, especially in the context of sensitive information management.

### 5.3 Redaction Challenges

Our analysis revealed widespread acknowledgment of challenges associated with document redaction processes, though only four of the 30 responding authorities provided comprehensive details about these difficulties. Three primary challenges emerged from the data:

*5.3.1 Resource Constraints and Inefficiencies*

Respondents consistently identified the time-intensive nature of manual redaction as a significant barrier, particularly when processing large or complex documents. The Northern Ireland Ambulance Service Health and Social Care Trust emphasized this challenge in their response: *"Time-Intensive Nature: Manual redaction is resource-intensive, particularly for large or complex documents."*

This resource constraint represents more than a mere operational inconvenience—it constitutes a fundamental limitation of current redaction approaches within the context of increasing transparency demands. The linear relationship between document volume and processing time creates an unsustainable model as Freedom of Information requests continue to rise. This bottleneck directly impacts statutory compliance, as authorities struggle to meet the 20-working-day response deadline mandated by the Freedom of Information Act when faced with document-heavy requests requiring extensive redaction. On the other hand, Bradford District Care NHS Foundation Trust elaborated on specific technical limitations, reporting frequent software failures during redaction processes (notably Adobe Acrobat crashes), difficulties handling voluminous documents, and the repetitive, labour-intensive nature of manual redactions. These technical failures compound efficiency problems, creating unpredictable workflows that further strain limited resources. The combination of software limitations and labour-intensive processes reveals a critical gap between the technological capabilities currently deployed and the operational demands placed on public authorities.

*5.3.2 Accuracy and Consistency Concerns*

The data revealed persistent concerns about human error in the redaction process, even among trained staff. As the Northern Ireland Ambulance Service noted: *"Human Error: Despite thorough training, occasional errors may occur when identifying sensitive information.* "This admission is particularly significant within the context of public authorities, where a single redaction error can lead to serious data protection breaches with legal, financial, and reputational consequences. The phrase "despite thorough training" suggests an implicit recognition that human error remains an inherent risk that cannot be entirely eliminated through conventional means. This creates a fundamental tension between the statutory obligation to protect sensitive information and the practical limitations of human-centred redaction processes.

Another case, including Bradford District Care NHS Foundation Trust, highlighted the elevated risk of inadvertently missing sensitive information and inconsistencies in redaction outcomes between different staff members. These inconsistencies point to a deeper systemic issue: the subjective nature of redaction decisions when left entirely to human judgment. Without technological standardization, the same document processed by different staff members may yield substantially different redaction outcomes, undermining the consistency that is critical for legal compliance and public trust. Similarly, Queen Mary University of London emphasized the challenge of maintaining consistency:

> The biggest challenge for human reviewers is ensuring that the requester of the information (in a subject access request) is provided with everything that we hold and that the redaction is consistent even when the volume of pages reviewed is very high.

The university's emphasis on "everything that we hold" reveals another dimension of the consistency challenge: comprehensiveness. As document volumes increase, the cognitive burden on human reviewers grows exponentially, making it increasingly difficult to maintain consistent decision-making across hundreds or thousands of pages. This observation underscores a fundamental scaling problem in current redaction approaches that directly impacts both transparency obligations and data protection compliance.

*5.3.3 Technical Limitations and Format Compatibility*

Respondents identified specific challenges related to document formats, particularly non-standard or handwritten materials. The Northern Ireland Ambulance Service noted: *"Complex Document Formats: Non-standard or handwritten formats can be more challenging to redact effectively.* "This observation highlights a critical technical gap in current redaction capabilities that extends beyond mere inconvenience. Public authorities must process a diverse array of document types—from handwritten notes and legacy paper records to complex modern digital formats—yet current redaction tools are primarily optimized for standardized digital text. This format incompatibility creates significant information governance risks, as non-standard documents may receive less thorough redaction or require disproportionate resources to process correctly. Moreover, this limitation reflects the technological inadequacy of current redaction approaches in handling the heterogeneous document ecosystems typical in public sector organizations.

Notably, even authorities that have not yet adopted AI-based tools recognized the potential benefits automation could offer. Queen Mary University of London acknowledged: *"This is something that could potentially be aided by AI-based redaction systems; however, at the moment, we do not use any, so we cannot comment on the challenges of using such systems.* "This statement reveals an important emerging awareness among public authorities regarding the limitations of current approaches and the potential technological solutions. The phrasing suggests a cautious openness to AI adoption rather than active resistance—the barrier appears to be one of implementation knowledge and resources rather than institutional reluctance. This receptivity is particularly significant given the otherwise conservative approach to technological change often observed in public sector information governance. The acknowledgment of AI's potential represents an important shift in institutional perspective that could facilitate future adoption if appropriate governance frameworks, training programs, and implementation support were made available.

## 6 DISCUSSION

### 6.1 Poor Recordkeeping

The findings of this study underscore the pressing need for improved recordkeeping and transparency in public-sector redaction processes while also extending prior research on document redaction to AI adoption. Drawing on our quantitative data, a pervasive lack of formal redaction policies and AI recordkeeping among UK public authorities emerges as a critical concern, with 50% of responding authorities reporting "information not held" regarding redaction procedures. This outcome aligns closely with Chen and Kirkham's [25] observation that 66.4% of surveyed authorities lacked official redaction procedures, suggesting that redaction governance across UK government bodies may be no more robust than previously reported.

From a legal standpoint, the absence of documentation in most cases is itself concerning. Public-sector organizations cannot credibly claim to be striving for compliance with the UK General Data Protection Regulation (GDPR) if they lack a documented redaction procedure or remain unable to locate the process intended for routine implementation. Moreover, the study's findings on limited recordkeeping surrounding AI usage are consistent with Roberts, Babuta, Morley, Thomas, Taddeo and Floridi [43], who highlight the United Kingdom's fragmented regulatory environment and the lack of a dedicated AI framework.

In contrast to the UK's fragmented approach, the U.S. National Archives and Records Administration [68] has developed structured guidance that may specifically address records management implications of AI systems, including documentation requirements for automated processing. While operating under different regulatory frameworks, this example demonstrates how dedicated governance can bridge the gap between technological capabilities and recordkeeping requirements in AI implementation. The absence of similar guidance in the UK context is particularly problematic since these regulatory uncertainties hinder many AI innovations from achieving scalability, leaving their broader impact in public-sector contexts largely unknown [69]. As a result, such consequences will be further amplified by the absence of key data protection measures, such as formally documented redaction protocols, thereby increasing both operational and legal risks [70].

### 6.2 The Need for Standardized Guidelines in Redaction Practices

Our analysis revealed a critical governance gap in redaction practices across UK public authorities. The absence of uniform, sector-wide standards has created a fragmented implementation landscape where many authorities resort to rudimentary methods and existing general guidelines rather than rigorous approaches. Specifically, at the regulatory level, the regulatory guidance from authoritative bodies like the National Archives fails to provide the prescriptive frameworks necessary for consistent implementation, creating a particularly problematic void for emerging AI-based redaction technologies. This standardization deficit directly impacts AI adoption in two significant ways: first, by generating regulatory uncertainty that discourages technological innovation, and second, by preventing the establishment of clear assessment criteria for evaluating automated redaction outputs. The resulting implementation hesitancy represents a missed opportunity to leverage technological advances that could address the inefficiencies identified in our findings. The consequences of this standardization vacuum extend beyond mere operational inefficiency—they create systemic vulnerabilities that have resulted in high-profile information governance failures. The AstraZeneca Contract controversy exemplifies how inadequate redaction governance translates into tangible risks, where sensitive contractual information was inadvertently disclosed despite redaction attempts [6]. Such incidents reveal how the current patchwork of inconsistent practices undermines both the efficacy of transparency initiatives and the integrity of data protection efforts, ultimately threatening public trust in information governance systems.

Therefore, this suggests an urgent need for coordinated regulatory action to establish technology-neutral redaction standards that can accommodate both current manual approaches and emerging AI-based methods. Also, we require the establishment of governance frameworks that delineate clear and consistent parameters for redaction practices. Such frameworks should not only articulate the processes for integrating AI systems but also specify mechanisms for human oversight and post-redaction review aimed at minimizing errors. In formulating these guidelines, it is imperative to ensure their alignment with existing data protection regulations, thus clarifying the legal responsibilities of authorities seeking to automate sensitive information management. Regulatory entities such as the ICO could play a pivotal role by expanding their AI-specific guidance and developing consistent protocols that facilitate broad-scale adoption of reliable redaction practices. Additionally, possible collaboration with cross-sectoral consortia, for example, through the Digital Regulation Cooperation Forum (DRCF), may help harness diverse expertise and streamline the formulation of best practices [71]. A standardized, transparent approach would likely enhance confidence in AI-mediated methods and mitigate inconsistencies arising from manual redaction processes. Ultimately, by promulgating a well-defined framework for redaction, public authorities would be better positioned to safeguard sensitive data, fulfil their obligations under FOI legislation, and advance the government's broader objective of exemplifying safe and ethical AI deployment in public services.

**6.3 Gaps in AI-Specific Training and Operational Risks**

The findings highlight a significant gap in the formal training on redaction practices among UK public authorities. Where training does exist, it is often informal or reactive, largely delivered through "in-house" or "on-the-job" methods. While such approaches can be flexible and cost-effective, they lack standardized curricula and may lead to uneven skill development. In fact, some authorities provided only minimal details about the nature and content of their training, suggesting a limited appreciation of the underlying complexities of the redaction process. Indeed, redaction can appear deceptively straightforward, an illusion that appears to have led many authorities to adopt a "light-touch," rather than a rigorous, approach—potentially exposing them to operational and legal risks. These challenges are amplified when AI-driven tools enter redaction workflows. Such systems not only demand proficiency in data protection legislation and information governance principles but also require specialized technical training so that staff can correctly operate, interpret, and critically evaluate algorithmic outputs.

To mitigate such potential risks, Figure 4 illustrates a two-stage redaction process that integrates manual training with AI-driven tools, underscoring the importance of both human oversight and technological automation in mitigating the risks of disclosing sensitive information. This approach addresses key issues observed in many public authorities, such as the lack of formal redaction training and the significant scalability challenges of manual processes, by first establishing a foundation of human competence before introducing AI systems. Once staff members have attained sufficient proficiency in manual redaction, AI-assisted workflows can be deployed to handle high-volume or routine tasks, with the system automatically assigning confidence scores to identify which documents or passages require human review [72]. This method not only reduces human workloads but also facilitates more effective error monitoring: medium- to low-confidence items are flagged for careful assessment, while documents deemed sufficiently clear by the AI proceed with minimal intervention [73]. Moreover, by ensuring alignment with regulations such as the UK GDPR and the EU Artificial Intelligence Act (AIA)—notably Article 14(1) on maintaining meaningful human oversight—the hybrid model can help public authorities navigate legal complexities while improving efficiency. Nevertheless, challenges such as calibrating thresholds for human review and addressing the potential biases embedded in AI algorithms remain crucial considerations. To mitigate these challenges, an escalation mechanism involving senior leadership serves as a safeguard for reconciling discrepancies and ensuring adherence to ethical standards. Ultimately, this two-stage process exemplifies how AI can bolster operational capacity in redaction without displacing the irreplaceable judgment and contextual expertise of human reviewers.

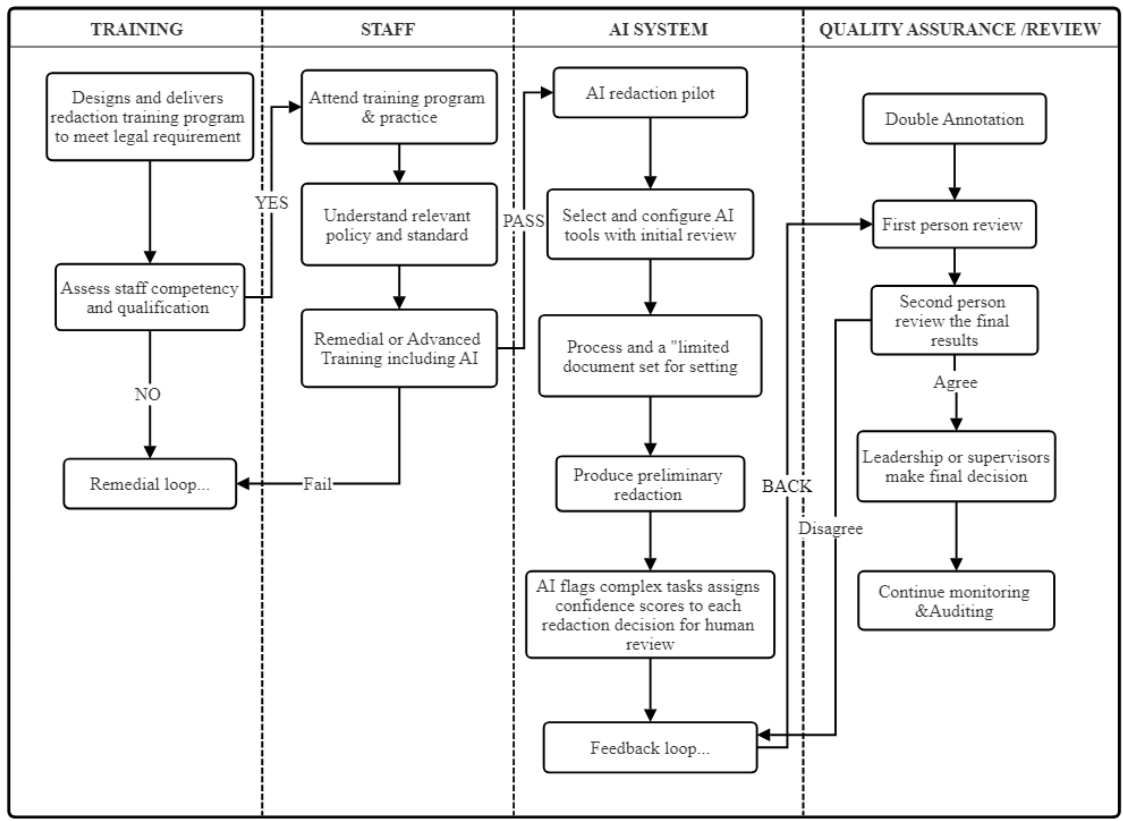

Figure 4: This figure illustrates a redaction training program is first designed to satisfy legal standards. Staff competency is assessed: those who pass advance to AI-assisted or advanced training; those who fail receive remedial support. Trained personnel then work with the AI redaction pilot, selecting and configuring tools, and processing a limited document set to refine settings. The system generates preliminary redactions, highlighting complex tasks with confidence scores for human review; feedback from this review loop continually refines AI accuracy. For quality assurance, two reviewers independently evaluate the redactions; if they agree, the final decision goes to leadership, while disagreements are resolved by supervisors. Ongoing monitoring and auditing further ensure continuous compliance and improvement.

## 7 LIMITATIONS AND FUTURE RESEARCH

This study has provided an initial snapshot of existing AI redaction practices within public authorities in the United Kingdom. Being an FOI study, the information we obtain is limited to recorded information and also heavily relies on public authorities complying with the legislation, which does not always happen. Future studies are worth conducting broader quantitative research covering a larger sample of public authorities, including a nationwide survey or a series of case studies across more agencies, which would offer more generalized data on AI adoption rates, redaction error frequencies and outcomes. Such studies could track, for instance, how many authorities have started piloting AI tools, what proportion of documents are being auto-redacted, and whether any measurable improvements (or failures) have occurred.

Also, large-scale longitudinal research would be valuable in observing whether the introduction of AI reduces the incidence of sensitive data leaks or speeds up FOI response times. This could involve analyzing FOI request processing metrics before and after AI implementation or comparing cohorts of organizations with and without AI assistance. By aggregating evidence from many sources, researchers can better assess the true impact of AI on transparency and privacy protection, informing both practice and policy at the national level. Another fruitful avenue is comparing AI redaction practices across different sectors or types of public bodies. The UK public sector is diverse – what works for a local council may differ for a central government department or a healthcare trust handling patient records. Future research might examine how AI adoption and redaction needs vary between local government, central government, law enforcement, healthcare, education, and so on. Additionally, international comparisons could provide insight: looking at how public authorities in other countries (with similar FOI laws or privacy regulations) are using AI for redaction could highlight best practices or alternative strategies. A cross-sectoral analysis can also identify unique challenges – for example, one sector might struggle more with specific types of data (like medical terminology in documents) that an AI needs to handle. Understanding these nuances will help tailor solutions to different contexts and encourage knowledge sharing between organizations. It may be useful to conduct workshops or collaborative studies where multiple authorities come together to pilot an AI tool and document their experiences across organizational boundaries.

Finally, given that AI-driven document redaction in the public sector is a relatively new and evolving area, future research should rigorously evaluate the performance of automated redaction tools. This entails developing benchmarks and test datasets representative of the documents handled under FOI or data protection requests. Researchers could measure how well various AI solutions identify and redact sensitive information, comparing their precision (avoiding false positives where non-sensitive data is blacked out) and recall (avoiding false negatives where something sensitive is left unredacted). Such evaluations might reveal, for example, how an AI performs on different document formats (scanned PDFs vs. digital text) or types of content (structured forms vs. free-text reports). By publishing clear metrics on accuracy, error rates, and processing speed, these studies would provide an evidence base those public authorities can use when considering tool adoption. Overall, by focusing on tool performance and human-AI interaction, research can drive technological refinements that increase confidence in automated redaction.

## 8 CONCLUSION

The result of this work suggests that most organizations lack standardized policies and relevant training, which exposes vulnerabilities in the transparency, accountability, and efficiency of the redaction process. Furthermore, the research underscores the need for incorporating AI tools with human oversight to mitigate the risks of error and bias in document redaction practices. The real-world impact of this research is substantial. By promoting the development of AI-driven redaction policies, we can ensure greater consistency and public trust. Also, the insights gained could inform the creation of legal and regulatory frameworks that balance technological innovation with the protection of individual rights. These findings contribute to the ongoing conversation around ethical data practices and offer actionable recommendations for policymakers to address current gaps in public sector information management. In the broader picture, this research fills the gap by providing empirical evidence about the implications of AI use for public governance. Future work could explore expanding the scope of the study to include a broader set of institutions or investigate the role of specific AI tools in redaction quality. This will not only improve transparency but also contribute to a more robust, ethical governance framework for the digital era.